\documentclass[]{spie}  %>>> use for US letter paper
\usepackage{amsmath,amsfonts,amssymb}
\usepackage{graphicx}
\usepackage[colorlinks=true, allcolors=blue]{hyperref}
\usepackage[super]{nth}
\usepackage{comment}
\usepackage{tabularx}
\usepackage{subcaption}
\title{Automated femur segmentation from computed tomography images using a deep neural network}

\author[a]{P.A. Bjornsson$^*$}
\author[b]{B. Helgason}
\author[c]{H. Palsson}
\author[d]{S. Sigurdsson}
\author[d,e]{V. Gudnason}
\author[a,f]{L.M. Ellingsen}
\affil[a]{Dept. of Electrical and Computer Engineering, The University of Iceland, Reykjavik, Iceland}
\affil[b]{Institute for Biomechanics, ETH Zürich, Zürich, Switzerland}
\affil[c]{Dept. of Industrial Engineering, Mechanical Engineering, and Computer Science, The University of Iceland, Reykjavik, Iceland}
\affil[d]{The Icelandic Heart Association, Kopavogur, Iceland}
\affil[e]{Dept. of Medicine, The University of Iceland, Reykjavik, Iceland}
\affil[f]{Dept. of Electrical and Computer Engineering, The Johns Hopkins University, Baltimore, MD 21218, USA}

\authorinfo{Further author information: (Send correspondence to P.A.B.)\\P.A.B.: E-mail: pab11@hi.is\\  L.M.E.: E-mail: lotta@hi.is}

\begin{document} 
\maketitle

\begin{abstract}
Osteoporosis is a common bone disease that occurs when the creation of new bone does not keep up with the loss of old bone, resulting in increased fracture risk. Adults over the age of 50 are especially at risk and see their quality of life diminished because of limited mobility, which can lead to isolation and depression. We are developing a robust screening method capable of identifying individuals predisposed to hip fracture to address this clinical challenge. The method uses finite element analysis and relies on segmented computed tomography (CT) images of the hip. Presently, the segmentation of the proximal femur requires manual input, which is a tedious task, prone to human error, and severely limits the practicality of the method in a clinical context. Here we present a novel approach for segmenting the proximal femur that uses a deep convolutional neural network to produce accurate, automated, robust, and fast segmentations of the femur from CT scans. The network architecture is based on the renowned u-net, which consists of a downsampling path to extract increasingly complex features of the input patch and an upsampling path to convert the acquired low resolution image into a high resolution one.  Skipped connections allow us to recover critical spatial information lost during downsampling. The model was trained on 30 manually segmented CT images and was evaluated on 200 ground truth manual segmentations. Our method delivers a mean Dice similarity coefficient (DSC) and \nth{95} percentile Hausdorff distance (HD95) of $0.990$ and $0.981$ mm, respectively. 
\end{abstract}


\section{Introduction}
A consequence of ageing societies is a higher prevalence of chronic diseases amongst older populations. In the coming years, this will lead to  shortages of qualified health care practitioners and mounting health care costs as a result of the added strain placed by these longevous populations. It appears that the current status quo of health care systems is unsustainable and must be restructured from face-to-face based care to a more decentralized, home based care that places emphasis on prevention rather than treatment. One disease that disproportionately affects those over the age of 50 is osteoporosis - a bone disease that occurs when the body loses too much bone, makes too little bone, or both, resulting in reduced bone mass, which leads to increased bone fragility and heightened fracture risk with age. Hip fractures are associated with some of the most dire socioeconomic consequences: those who incur a fracture typically experience a steep decline in physical, mental, and emotional function and, in 50-55\% of cases, individuals are left with residual walking disability and in 15-30\% of cases, these individuals must be remanded to institutional care \cite{ekstrom,carpenter,magaziner,neval,osnes}. What is most startling is that 11-23\% of individuals will be deceased six months after incurring the fracture, increasing to 22-29\% after one year has passed since the incident \cite{haleem}. 

Subject-specific, image-based finite element (FE) analysis of bone is a popular approach in biomechanics that has gained considerable traction in recent years \cite{fe_grein}. By exploiting the gray-scale features of the CT images in concert with the respective segmentation mask, a screening tool for hip fracture risk prediction can be brought to fruition. Such a clinical screening tool could prevent potentially devastating fractures for patients, as well as to dramatically relieve the economic burden of hip fractures on our healthcare systems. However, these methods have yet to be implemented for clinical use, since the segmentation of the proximal femur currently requires manual input, which severely limits their use in clinical context.

%Numerous methods for proximal femur segmentation still require a ``user-in-the-loop'' paradigm to manually correct unsatisfactory segmentations and produce acceptable masks for FE modeling. These methods suffer from an inability to process large cohorts of data, and such a lack of robustness renders clinical application impractical. 
A number of methods have been implemented to surmount the problem of bone segmentation, ranging from thresholding techniques to graph-cut methods \cite{kim, chang, younes, yves}. Current segmentation methods often require a ``user-in-the-loop" paradigm in order to manually correct segmentations to produce acceptable masks for FE modeling and/or their processing time is too long for clinical use. This lack of robustness is costly in terms of time and the need for a highly trained specialist to manually correct the segmentations. Consequently, these methods cannot process larger cohorts to the same degree as fully automated ones.
In recent years, the application of deep neural networks (DNNs) in image segmentation has gained considerable attention. The use of deep neural networks as a viable option for biomedical image segmentation is a direct consequence of the u-net architecture proposed by Ronneberger et al. \cite{unet}. In this impactful paper, they demonstrated how the u-net architecture can produce fast and precise segmentations without relying on a large training set. However, the process by which ground truth segmentation masks are authored is typically manual delineation, which is a taxing and time-consuming process that is ideally only implemented for a small training data set. Despite the need for ground truth labels, deep neural networks have become the state-of-the-art approach in medical imaging \cite{shao, huo}. A segmentation prediction on novel data can be generated autonomously (i.e., without human intervention) in a matter of seconds, which is often an order of magnitude faster than preceding methods that do not make use of DNN's. 
%Three-dimensional convolutional neural networks (CNNs) have, as a result, become the state-of-the-art method in biomedical imaging. 
Zhao et al. \cite{zhao} proposed an automated, patch-based three-dimensional v-net architecture \cite{v-net} and reported a mean DSC of $0.9815\pm 0.0009$. Another successful method based on the u-net architecture \cite{unet} was conducted by Chen et al. \cite{chen}, who reported a mean DSC of $0.9688\pm 0.0095$. The major drawback to these two methods, however, is their dependence on large training sets that require ground truth segmentations, as well as the small validation sets they have been evaluated on. The arduous nature of authoring these ground truth segmentations, as previously mentioned, is oftentimes not feasible and severely limits the practicality of these methods.

Here we propose a fully automated method for segmenting the proximal femur from CT scans that is accurate, robust, and faster than other state-of-the-art methods. We evaluate our method on 200 manually delineated ground truth annotations  from the Age, Gene/Environment Susceptibility-Reykjavik Study (AGES-RS)\cite{ages}, a unique longitudinal study of the Icelandic elderly, to demonstrate the accuracy and robustness of our proposed method.
\section{Methods}
\label{sec:methods}  % \label{} allows reference to this section
\subsection{Preprocessing and data augmentation}
The Icelandic Heart Association (IHA) provided us with CT images from the AGES-RS \cite{ages}: a cohort that consists of both men and women born between 1907 and 1935, monitored in Iceland by the IHA since 1967. This unique database of high-quality CT images contains roughly 4800 density calibrated CT scans of the proximal femur at baseline and 3300 scans of the same individuals acquired at a five-year follow-up. The in-plane resolution of each scan is $512\times512$ voxels with 88-178 slices and $0.977 \times 0.977 \times 1~\mathrm{mm}^3$ voxel size. 
The preprocessing of the data consisted of three steps: Firstly, each of the 30 CT images used for training was split in half and the left sides were subsequently mirrored to the right, resulting in 60 images of the right side hip/upper leg. Second, min-max normalization was used to linearly map the intensity values from Hounsfield units to the range $[0,1]$. Lastly, Otsu thresholding \cite{otsu-original} was implemented to identify the background voxels and allow for automatic cropping of the 3D images to decrease their size and therefore speed up the execution of the neural network. Since each extra voxel is computationally costly, the images were all cropped to eliminate unnecessary background voxels. 

\begin{figure*}[t]
\centerline{\includegraphics[width=\textwidth]{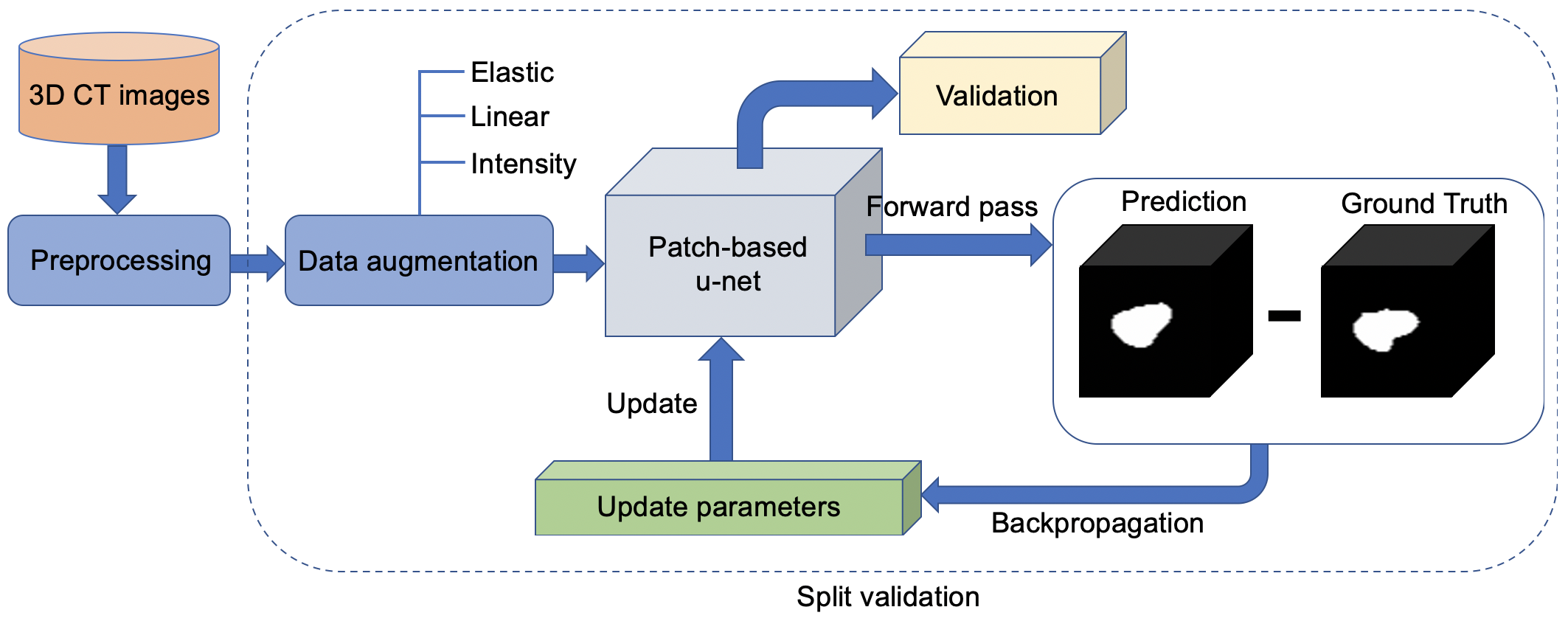}}
\caption{A flowchart showing the workflow of our proposed method.}
\label{workflow}
\end{figure*}

The proposed fully automated proximal femur segmentation pipeline is illustrated in Figure~\ref{workflow}. During training, a training/validation set of 30 CT images (i.e., 60 proximal femurs) from the AGES cohort \cite{ages} was fed into a neural network that exploited on-the-fly data augmentation on randomly cropped patches from a randomly selected batch of two images. On-the-fly data augmentation eliminates the need for excessive storage of augmented images by performing the augmentation prior to each optimization iteration. 
Data augmentation is imperative for maximizing the efficiency of a training set, since obtaining manual segmentations (i.e., ground truth segmentations) is an arduous, time-consuming process that does not cater to large cohorts. Hence, data augmentation is applied to teach the model invariance and robustness properties when only a limited data set is available. In addition, artificially expanding the training set with data augmentation has a regularizing effect, which avoids overfitting the model to a small subset of data. Deformations that capture variation within the data set can be simulated efficiently and assist the model in learning invariance between images \cite{unet}. Here we applied elastic deformations as well as linear-spatial and intensity transformations (i.e., scaling, rotation and brightness) to simulate the variability between patients' scans.
%using the Batchgenerators package \cite{batchgenerators} within the Medical Image Segmentation with Convolutional Neural Networks (MIScnn) framework\footnote{MIScnn is an open-source Python library and intuitive API for medical image segmentation pipelines \cite{miscnn}.}. 
Data augmentation, with random transformation parameters from pre-defined ranges, was implemented on-the-fly for each image before it was forwarded into the neural network. Each transformation had a 35\% likelihood of being applied to the image at hand, allowing the model to encounter a diverse set of images, thereby decreasing redundancy. The scaling, rotation, and brightness transforms were defined in the ranges $(0.95, 1.05)$, $(-3^{\circ}, 3^{\circ})$, and $(0.75,1.25)$, respectively, while the elastic deformation transform was tuned to $\alpha=(0,100)$ and $\sigma=(9,13)$.\footnote{Here, $\alpha$ denotes the scaling factor (controls the deformation intensity) and $\sigma$ denotes the smoothing factor (controls the displacement field smoothing) for the elastic deformation.}
On-the-fly data augmentation, in concert with parameter sharing \cite{lecun-boser} and batch normalization \cite{BN}, rendered the use of explicit regularization techniques unnecessary and even counterproductive. 

\subsection{The segmentation workflow and model architecture}
\begin{figure}[t]%
\centering
\includegraphics[width=\textwidth]{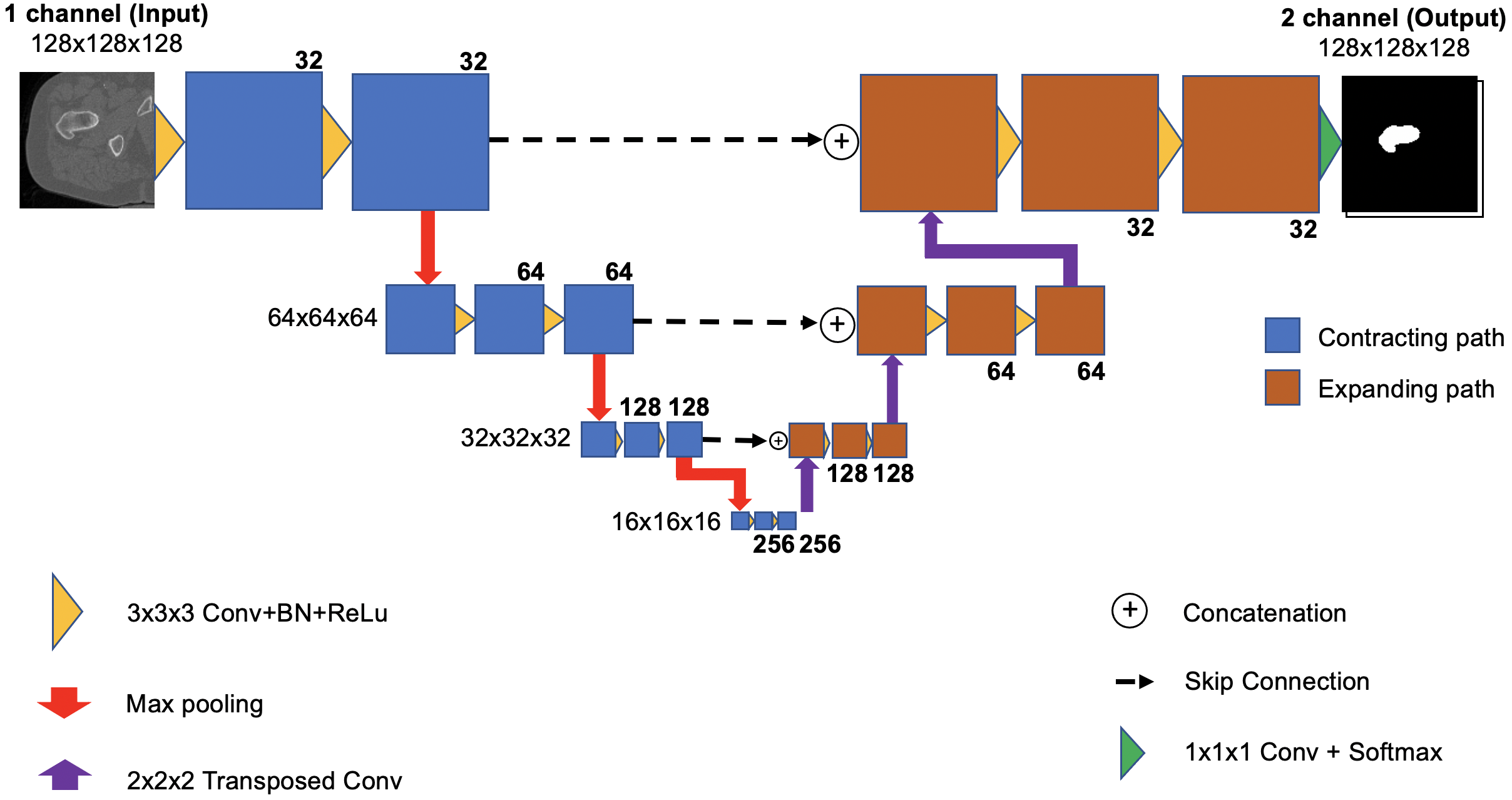}
\caption{A schematic of our proposed 3D u-net model. The bold numbers at the corners represent the number of feature maps (channels) per layer. For the sake of simplicity, feature maps are depicted in two dimensions.}
\label{unet_standard_schematic}
\end{figure}

As previously stated, the u-net architecture addresses two main issues: namely, the ability to train a model from a very small data set and the ability to produce precise segmentations despite the former. A schematic of the proposed architecture is shown in Figure~\ref{unet_standard_schematic}. The two paths that comprise the u-net structure are the contracting (downsampling) path and the expanding (upsampling) path. The former is the encoder and captures context with the use of stacked convolutional units and max pooling layers, while the latter is the decoder and allows for precise localization with the use of transposed convolutions.
In the final layer of the network, a $1\times1\times1$ convolution is used to map the feature map to the number of classes. By applying a softmax layer voxel-wise, a probabilistic map is outputted that gives each voxel a value in the range $[0,1]$; voxels with a probability $>0.5$ belong to the foreground class (proximal femur) and the rest to the background class. 
The proposed method was implemented using the MIScnn framework \cite{miscnn} in Python.

A patch-based model was adopted in consideration of memory constraints; each volume patch was $128\times128\times128$ voxels with an overlap of $64\times64\times64$ voxels. This patch size captures the entire femoral head, which is the most dynamic section of the proximal femur, and downsamples nicely, meaning that after each use of max pooling we are left with integer values for patch dimensions. During training, volume patches were randomly selected from the image volume and subsequently fed into the neural network. As the image patch is downsampled in the contracting path, the number of channels increases, which allows for the extraction of more complex features. Skip connections are used to pass features from the contracting path to the expanding path in order to recover spatial information lost during downsampling. At each step in the decoder, a skip connection allows for the concatenation of the output of the transposed convolutional layer with the corresponding feature map from the encoder. A convolutional layer is subsequently able to produce a more precise output based on this information.
The Dice loss function, defined as $1-\mathrm{DSC}$, quantified the performance of the model at each iteration of each epoch in order for the gradient descent algorithm to adjust the parameters during backpropagation. The validation set allowed us to gain insight into the performance of the model on unseen data so that hyperparameters could be adjusted accordingly.

Our model was trained on images from the AGES data set, comprising 260 proximal femurs with corresponding manually delineated segmentations, generated with a semi-automated delineation protocol, that served as ground truth annotations. A total of 54 images were used for training, 6 were used for validation, and the remaining 200 were used for evaluation. 
 The Adam optimizer ($\beta_1 = 0.9$ and $\beta_2 = 0.999$) was used to update the weights with an initial learning rate of $1 \cdot 10^{-4}$. Our model was trained using dual Nvidia GeForce GTX 1080 Ti GPUs for 300 epochs, which took
roughly 11.5 hours.

\section{Experiments and Results}
\label{results}
\begin{figure}[t!]
\centering
\includegraphics[scale=.43]{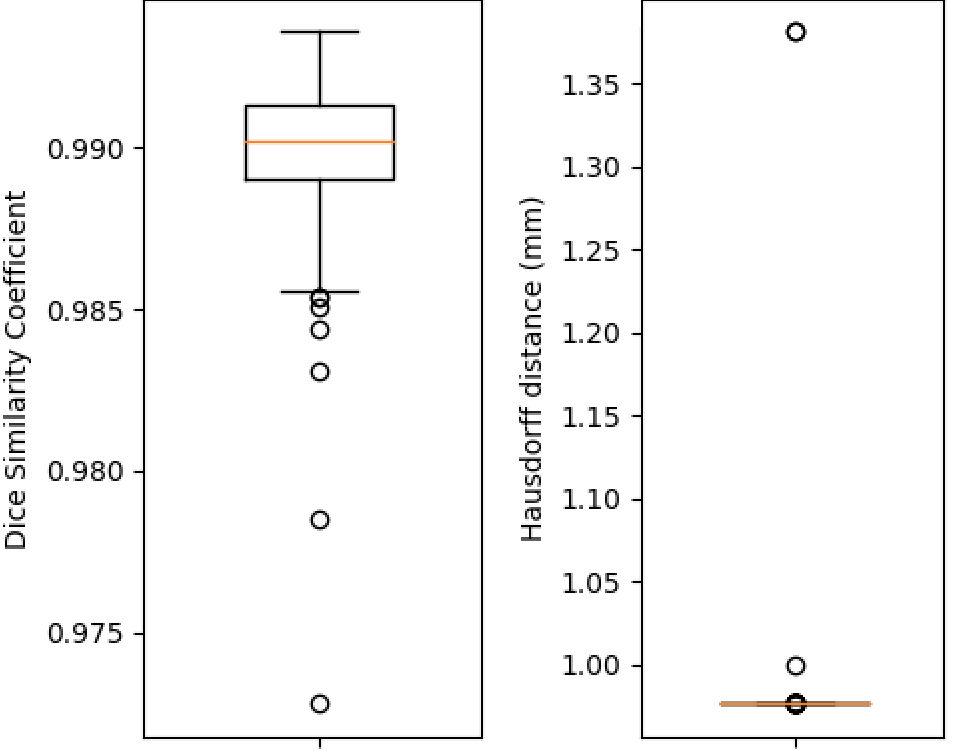}
\caption{The box plots above show the mean DSC (left) and the mean HD95 (right) for 200 proximal femurs.}
\label{boxplot_dice}
\end{figure}

\begin{figure}[t!]
\centering
\includegraphics[scale=.6]{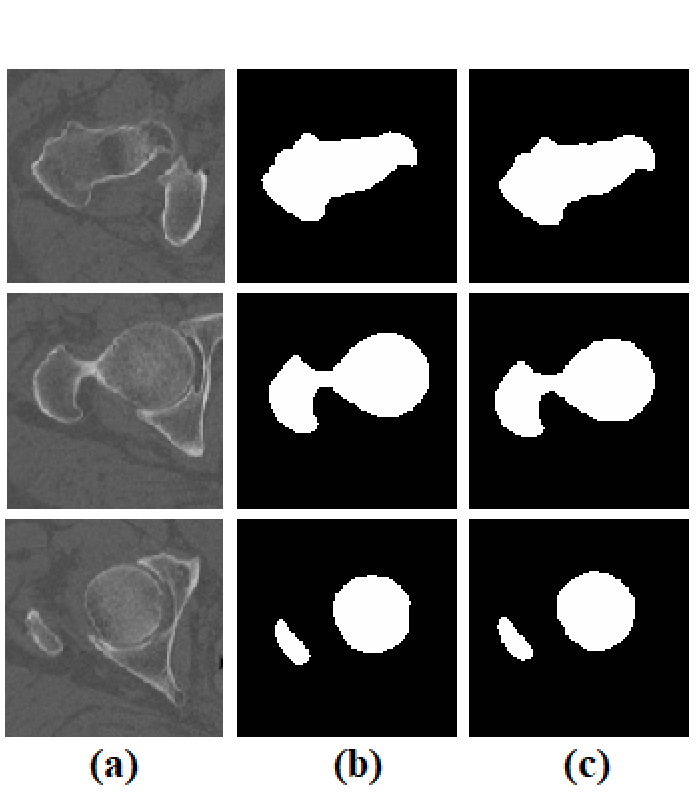}
\caption{A visual comparison of our segmentation results and the manual segmentation. \textbf{(a)} The original CT scans, \textbf{(b)} the segmentation predictions using the proposed method, \textbf{(c)} the ground truth segmentations.}
\label{boxplot_slice}
\end{figure}

% \subsection{Evaluation Criteria}
We evaluated the performance of our model on the aforementioned AGES data set by processing 200 proximal femurs that had been manually segmented (the current gold standard). We used the DSC and HD95 to evaluate the accuracy and robustness of our method. The former metric measures the overlap between the automatically generated segmentation mask and the ground truth manual mask. The HD95 metric, on the other hand, quantifies the largest segmentation error between two images and provides valuable insight into the performance of the model. The mean DSC on this evaluation set was $0.990 \pm 0.002$ and the mean HD95 was $0.981 \pm 0.040$ mm (see Fig. \ref{boxplot_dice}), demonstrating both high accuracy and robustness of the proposed method. The time for each segmentation prediction was 12-15 seconds, making the method viable for use in both large studies and clinical settings.
Figure \ref{boxplot_slice} displays a visual comparison between the output of our proposed method and the ground truth segmentation. Despite the thin and diffusive bone structure boundaries between the femoral head and the acetabulum (i.e., the bottom two rows of Figure \ref{boxplot_slice}), our method accurately labels the voxels that belong to the foreground class.

\section{Discussion and Conclusion}
This paper introduces a fully automated, accurate, fast, and robust method for proximal femur segmentation that produces segmentation results with sub-millimeter accuracy.
Our model addresses the biggest hurdles that have impeded prior methods. Firstly, it is fully automated and does not require a trained operator to make ad hoc corrections to unacceptable segmentation predictions. Secondly, the time our model takes to output a segmentation prediction (around 12-15 seconds per prediction) is well within reasonable bounds for clinical viability and can additionally be implemented to process large cohorts. 
The mean DSC was $0.990\pm0.002$ and mean HD95 was $0.981\pm0.040$ mm when evaluated on 200 manually segmented femurs. The proposed method is superior to preceding methods in terms of previously reported numbers of DSC and HD95 metrics and does not require any manual interaction. In the near future, we will conduct a more extensive evaluation on a larger cohort and, in turn, integrate the method into our existing FE pipeline, bringing it one step closer to becoming a clinically viable option for screening at-risk patients for hip fracture susceptibility.

\section{Acknowledgements}
This work was supported by the RANNIS Icelandic Student Innovation Fund.

\bibliography{report} % bibliography data in report.bib
\bibliographystyle{spiebib} % makes bibtex use spiebib.bst

\end{document}